\documentclass[12pt]{article}

\def\fun#1#2{\lower3.6pt\vbox{\baselineskip0pt\lineskip.9pt
\ialign{$\mathsurround=0pt#1\hfil##\hfil$\crcr#2\crcr\sim\crcr}}}

\def\be{\begin{equation}}
\def\ee{\end{equation}}
\def\beq{\begin{equation}}
\def\eeq{\end{equation}}
\def\bea{\begin{eqnarray}}
\def\eea{\end{eqnarray}}

\def\fun#1#2{\lower3.6pt\vbox{\baselineskip0pt\lineskip.9pt
\ialign{$\mathsurround=0pt#1\hfil##\hfil$\crcr#2\crcr\sim\crcr}}}

\def\a{\alpha}

\def\a{\alpha}
\def\b{\beta}

\def\be{\begin{equation}}
\def\ee{\end{equation}}
\def\beq{\begin{equation}}
\def\eeq{\end{equation}}
\def\bea{\begin{eqnarray}}
\def\eea{\end{eqnarray}}

\begin{document}

\title{Finite size giant magnon}

\author{Bojan Ramadanovic and Gordon W. Semenoff \\~~
\\{\small \it Department of Physics and Astronomy, University of British Columbia}\\
{\small\it Vancouver, British Columbia, Canada V6T 1Z1}}

\maketitle

\begin{abstract}
The quantization of the giant magnon away from the infinite size limit
is discussed.  We argue that this quantization inevitably leads to
string theory on a $Z_M$-orbifold of $S^5$. This is shown explicitly
and examined in detail in the near plane-wave limit.
\end{abstract}

\vskip 2cm

A significant amount of work on the AdS/CFT correspondence
\cite{Maldacena:1997re}-\cite{Witten:1998qj}
has been inspired the idea that the planar limit of ${\cal N}=4$
Yang-Mills theory and its string dual might be integrable models
which would be completely solvable using a Bethe
Ans\"atz~\cite{Minahan:2002ve}-\cite{Beisert:2003yb}.
Computation of the conformal dimensions of composite operators in
${\cal N}=4$ Yang-Mills theory can be mapped onto the problem of
solving an $SU(2,2|4)$ spin chain.  It is known that the spin chain
simplifies considerably in the limit of infinite length where
dynamics are encoded in the scattering of magnons and integrability
would imply a factorized S-matrix~\cite{Arutyunov:2004vx}. Beginning
with this limit, a strategy advocated by Staudacher
\cite{Staudacher:2004tk}, Beisert showed that a residual $SU(2|2)^2$
supersymmetry and integrability determine the ${\cal N}=4~$ S-matrix
up to a phase~\cite{Beisert:2005tm},\cite{Beisert:2006qh}.  More
recent work constrains \cite{Janik:2006dc} and essentially computes
this phase \cite{Hernandez:2006tk}-\cite{Benna:2006nd}.

An important problem that the integrability program would eventually
have to address is that of finite size corrections. In fact, recent
four-loop computations of short operators
\cite{Fiamberti:2007rj},\cite{Keeler:2008ce} suggest that the most
advanced form of the integrability Ans\"atz,  due to Beisert, Eden
and Staudacher \cite{Beisert:2006ez}, is likely valid only in the
infinite size limit and it is spoiled by finite size effects. In the
gauge theory, these effects are thought to stem from wrapping
interactions~\cite{Ambjorn:2005wa}-\cite{Janik:2007wt}.

A place where some progress has been made in studying finite size
effects is the spectrum of a single magnon. The Bethe Ans\"atz
implies that the energy spectrum of a single magnon (at least in
infinite volume) has the form
\begin{equation}\label{spectrum}
\Delta-J = \sqrt{1+\frac{\lambda}{\pi^2}\sin^2\frac{p_{\rm mag}}{2}}
\end{equation}
Here, $\Delta$ is the conformal dimension and $J$ is a $U(1)\in
SU(4)$ R-charge which also dictates the length of the spin chain.
$p_{\rm mag}$ is the magnon momentum.  The string theory dual of the
magnon of the infinite size system, the ``giant magnon'', was
identified by Hofman and Maldacena \cite{Hofman:2006xt} who showed
that it had energy spectrum the expected leading large $\lambda$
limit of (\ref{spectrum}). Then it was noted that the giant magnon
solution could also be found in finite volume
\cite{Arutyunov:2006gs}-\cite{Klose:2008rx} where an asymptotic
expansion of its spectrum is
\begin{equation}\label{spectrum2}
\Delta-J = \frac{\sqrt{\lambda}}{\pi}\left|\sin\frac{p_{\rm mag}}{2}\right|
-4\frac{\sqrt{\lambda}}{\pi}\left|\sin\frac{p_{\rm mag}}{2}\right|^3
e^{-2-J\pi/\sqrt{\lambda}|\sin{\tiny\frac{p_{\rm mag}}{2}}|}+\ldots
\end{equation}
The finite size corrections are exponentially small with large $J$.
This was first found in the beautiful paper by Arutyunov, Frolov and
Zamaklar \cite{Arutyunov:2006gs} where, in their approach, it was
noted that the spectrum depended on a light-cone gauge fixing
parameter. This was not a problem in the strict infinite volume limit,
which turned out to be gauge invariant, but it afflicted the
exponentially small corrections. Arutyunov et.al.~attributed this
gauge variance to the fact that a single magnon with non-zero momentum
$p_{\rm mag}$ is not a physical state of either the string theory or
its dual, ${\cal N}=4$ Yang-Mills theory. In the gauge theory, a
single magnon would be an excitation of the exactly known
ferromagnetic ground state of the spin chain, the
${\tiny\frac{1}{2}}$-BPS chiral primary operator ${\rm Tr} Z^J$, which
has exact conformal dimension the classical value, $\Delta=J$,
protected by supersymmetry. We shall denote the three complex scalar
fields of ${\cal N}=4$ Yang-Mills theory as $(Z,\Psi,\Phi)$.  The
sixteen states of the Magnon multiplet are obtained by a spin flip --
a single insertion of $D_\mu Z$ or another scalar or a fermion into
the trace.  They form a short multiplet of the $SU(2|2)\times SU(2|2)$
subalgebra of $SU(2,2|4)$ which commutes with ${\rm Tr }Z^J$.  Because
of cyclicity of the trace, all positions where one could flip one spin
in the ground state of the spin chain are equivalent and $\sum_n
e^{inp}{\rm Tr}Z^n\Psi Z^{J-n}\sim \delta(p)$, the magnon momentum
must vanish\footnote{ This was not a problem in the limit of the
infinite chain discussed in Ref.~\cite{Hofman:2006xt} since there
could be other operators present to block cyclicity of the trace and
they could be placed infinitely far away from the magnon so that any
wave-packet state of the magnon is isolated. }.  Single magnon states
with finite magnon momentum do not exist.\footnote{ We also note that
these states can be obtained by commutators of symmetry generators and
${\rm Tr}Z^J$ so they are all in the same ${\tiny\frac{1}{2}}$-BPS
multiplet of the full $SU(2,2|4)$ algebra and have exact conformal
dimensions $\Delta-J=1$ which agrees with (\ref{spectrum}) and
(\ref{spectrum2}) when $p=0$.}

The Hofman-Maldacena giant magnon \cite{Hofman:2006xt} is a soliton
solution of the bosonic part of the IIB sigma model propagating on an
$R^1\times S^2$ subspace of $AdS_5\times S^5$.  They showed that a
magnon corresponds to a closed string with an
open boundary condition, where the azimuth angle spanned by the two
ends of the string corresponds to $p_{\rm
mag}$. Ref.~\cite{Arutyunov:2006gs} argued that the open boundary
condition led to a modification of the level-matching condition and
gauge parameter dependence of the spectrum was a result.
In Ref.~\cite{Astolfi:2007uz} it was suggested that the single
magnon is well-defined as the twisted state of a closed string on an
orbifold -- where the orbifold group acts in such a way that it
identifies the ends of the string, resulting in a legitimate state
of closed string theory.  This was advocated as a way to study the
spectrum of a single magnon in a setting where it is a physical
state and there are no issues with gauge invariance. The giant
magnon spectrum was computed there and an asymptotic expansion in
the size of the system yields (\ref{spectrum2})  (all results in
this picture are identical to what Ref.~\cite{Arutyunov:2006gs}
obtain if their gauge parameter $a$ is set to zero). In the
following we shall develop this idea further. {\it Our main
observation will be that if we consider the single magnon state in
the IIB string theory with the boundary condition that the string is
open in the direction of magnon motion, we are inevitably led to an
orbifold.}

 To get the gist of our argument, consider the following
(drastically oversimplified) example of the closed bosonic string on
flat Minkowski spacetime where we legislate that one of the string
coordinates is not periodic, but obeys the ``magnon'' boundary
condition $ X^1(\tau,\sigma=2\pi) = X^1(\tau,0)+p_{\rm mag}$ and all
other variables, including $ \partial_\sigma X^1(\tau,\sigma)$ are
periodic.  Then, a solution of the worldsheet equation of motion
$\left(\partial_\tau^2-\partial_\sigma^2\right)X^1=0$ with the
appropriate boundary condition is \cite{Polchinski:1998rq} $ X^1=
x^1+ \alpha' p^1\tau + \frac{\sigma}{2\pi} p_{\rm mag} + {\rm
~oscillators} $. One of the Virasoro constraints is the level
matching condition $ L_0 - \tilde L_0 = 0$ which takes the form
\begin{equation} N-\tilde N +p^1\frac{p_{\rm mag}}{2\pi}
=0\label{level}\end{equation} where $ N=\sum_{n=1}^\infty
\alpha_{-n}\cdot\alpha_n$ and $\tilde N=\sum_{n=1}^\infty \tilde
\alpha_{-n}\cdot\tilde \alpha_n $. Since the spectra of the
operators $N$ and $\tilde N$ are integers, there is no solution of
the level-matching condition unless $p^1p_{\rm mag} =2\pi\cdot{\rm
integer}$, i.e.~the momentum $p^1$ is quantized in units of
integer$\cdot 2\pi/p_{\rm mag}$. This is identical to (and
indistinguishable from) the situation where the dimension $X^1$ is
compactified with radius $R= \frac{p_{\rm mag}}{\rm integer}$ and
where we consider a wrapped string with fixed momentum which is then
quantized in units of $\frac{2\pi}{R}$. We see that the magnon
boundary condition leads us to string theory on a simple orbifold, a
periodic identification of the direction in which the magnon
boundary condition was taken. We shall observe a similar fact for
the more complicated case of a single magnon on $AdS_5\times S^5$
background.


The bosonic part of the IIB sigma model on
$AdS_5\times S^5$ and in the conformal gauge is
\begin{eqnarray}
{\cal L}=-\frac{\sqrt{\lambda}}{4\pi} \left\{ -\left( \frac{
1+\frac{Z^2}{4}}{1-\frac{Z^2}{4}}\right)^2\partial_aT\partial^aT
+\left( \frac{
1}{1-\frac{Z^2}{4}}\right)^2\partial_aZ\cdot\partial^aZ \right.
\nonumber \\ \left. +\left( \frac{
1-\frac{Y^2}{4}}{1+\frac{Y^2}{4}}\right)^2
\partial_a\chi\partial^a\chi+
\left( \frac{ 1}{1+\frac{Y^2}{4}}\right)^2
\partial_aY\cdot\partial^aY\right\}
\label{sigmamodelaction}
\end{eqnarray}
supplemented by Virasoro constraints.  The eight fields $\vec Z$ and
$\vec Y$ transform as 4-vectors under $SO(4)\times SO(4)\sim
SU(2)^4$. We will impose the magnon boundary condition on the angle
coordinate
\begin{equation}\label{chiboundarydoncition}
\chi(\tau,\sigma=2\pi)=\chi(\tau,\sigma=0)+p_{\rm mag}
\end{equation}
If $\chi(\tau,\sigma)=\tilde\chi(\tau,\sigma)+p_{\rm
mag}\sigma/2\pi$ with $\tilde\chi$ periodic,
\begin{eqnarray}\label{shiftedlagrangian}
{\cal L}[T,\vec Z,\chi,\vec Y]={\cal L}[T,\vec Z,\tilde\chi,\vec Y]
-\frac{\sqrt{\lambda}}{4\pi } \left(\left(\frac{p_{\rm
mag}}{2\pi}\right)^2+ \frac{ p_{\rm
mag}}{\pi}\tilde\chi'\right)\left( \frac{
1-\frac{Y^2}{4}}{1+\frac{Y^2}{4}}\right)^2
\end{eqnarray}
The effect of the magnon boundary condition is to add terms to the
action. These, as well as similar terms which appear in the Virasoro
constraints, will break some of the (super-)symmetries of the
background. The last terms in (\ref{shiftedlagrangian}) has the
 symmetries $SU(2)^2\times SU(2)^2\times R^2$ where the $R^2$ are
translations of $T$ and $\tilde\chi$.
 The bosonic part of the level-matching condition is
\begin{eqnarray}
0= \int_0^{2\pi}d\sigma\left\{\Pi_T T' +\Pi_Z Z' +
\Pi_{\tilde\chi}\tilde\chi' + \Pi_YY'\right\} +\frac{p_{\rm
mag}}{2\pi} J \label{levelmatching}
\end{eqnarray}
where $\Pi_\mu\equiv\partial{\cal L}/\partial\dot X^\mu$ are the
canonical momenta conjugate to coordinates $X^\mu$ and the charge $J
$ is the generator of translations of $\tilde \chi$,
$\chi\to\chi+{\rm const.}$
\begin{equation}\label{J}
J=\int_0^{2\pi}d\sigma\Pi_{\tilde\chi}  =\frac{\sqrt{\lambda}}{2\pi}
\int_0^{2\pi} d\sigma \left( \frac{
1-\frac{Y^2}{4}}{1+\frac{Y^2}{4}}\right)^2 \dot{\tilde\chi}
\end{equation}
Since $\chi\sim\chi+2\pi$, the eigenvalues of $J$ must be
integers.\footnote{When fermions are included, they could be
half-integers.} Furthermore, being generators of translations of the
worldsheet $\sigma$-argument of the fields, and the fields involved
being periodic in $\sigma$, the first four terms in
(\ref{levelmatching}) must be integers plus a possible
constant.\footnote{Consider the operator $\xi$ which has the property
$\left[ \xi,
\varphi(\sigma)\right]=i\frac{d}{d\sigma}\varphi(\sigma)$. Consider
eigenstates $|\alpha>$ and $|\alpha'>$ where $\xi
|\alpha>=\alpha|\alpha>$. If
$<\alpha'|\varphi(\sigma)|\alpha>=<\alpha'|e^{- i\sigma
\xi}\varphi(0)e^{ i
\sigma\xi}|\alpha>=e^{i(\alpha-\alpha')\sigma}<\alpha'| \varphi(0)
|\alpha>$, the matrix element obeys
$<\alpha'|\varphi(\sigma)|\alpha>=<\alpha'|\varphi(\sigma+2\pi)|\alpha>$
only when $\alpha-\alpha`={\rm integers}$.  The eigenvalues are equal
to integers plus a constant which is common to all eigenvalues.  If,
there is a reflection symmetry $\sigma\to2\pi-\sigma$ under which
$\xi\to-\xi$, the constant must be either an integer or
half-integer. }  Since the theory has a symmetry under $\sigma\to
2\pi-\sigma$, the constant must be either zero or one-half. Thus, the
spectrum of the first terms in (\ref{levelmatching}) is either
integers or integers$+{\tiny\frac{1}{2}}$.  To eliminate the second
possibility, we shall see that, in the plane wave limit, we can solve
for the spectrum explicitly and there we find that it is
integers. Then, since the spectrum should not change discontinuously
as the plane wave limit is taken, we conclude that it should always be
integers.

Since $J$ comes in units of integers, and the first four terms in
(\ref{levelmatching}) are integers, (\ref{levelmatching}) will only
have a solution if $\frac{p_{\rm mag}}{2\pi}$ is a rational number,
$\frac{m}{M}$.  Then, $J$ is quantized in units of $M$. This is
identical to what should occur for a m-times wrapped string on a
$Z_M$ orbifold of $AdS_5\times S^5$ where the orbifold group $Z_M$
makes the identification $\chi\to\chi+2\pi\frac{m}{M}$.

To get the superstring, we must include the fermions. For this, we
must decide what their boundary conditions will be.  It is clear
that, at large $J$, we will obtain the correct magnon supermultiplet
if we add them in such a way that, in the modification of the
Virasoro constraint (\ref{levelmatching}), $J$ also contains the
appropriate fermionic contribution $J\to\tilde J=\int
(\Pi_{\tilde\chi}\tilde\chi'+\Pi_\psi\Sigma\psi')$. This gives the
magnon boundary condition for the fermions
\begin{equation}\label{fermionboundarycondition}
 \psi(\tau,\sigma=2\pi)=e^{ip_{\rm mag}\tilde
J}\psi(\tau,\sigma=0)e^{-ip_{\rm mag}\tilde J} =e^{ip_{\rm
mag}\Sigma}\psi(\tau,\sigma=0)
\end{equation}
where $\Sigma= {\rm
diag}\left(\frac{1}{2}.-\frac{1}{2},\frac{1}{2},-\frac{1}{2}\right)$
and the orbifold identification is
\begin{equation}\label{orbifold1}
\left(\chi,\psi\right)\sim\left(\chi+p_{\rm mag},
e^{ip_{\rm mag}\Sigma}\psi\right)
\end{equation}
 All of the fermions have a twist in their boundary condition.  With
this identification, all supercharges transform non-trivially under
the orbifold group and all of the supersymmetries will be broken (in
fact, the supercharges are set to zero by the obtifold projection).
This twist in the fermion boundary condition and concomitant breaking
of supersymmetry is well known from orbifold constructions in string
theory \cite{Douglas:1996sw} and was outlined in detail in a context
similar to ours in Ref.~\cite{Alday:2005ww}.

Some supersymmetry can be saved if we impose a slightly more elaborate
identification:
\begin{equation}\label{orbifold2}
\left(\chi,Y_1+iY_2,\psi\right)\sim\left(\chi+p_{\rm mag}, e^{-ip_{\rm
mag}}(Y_1+iY_2), e^{ip_{\rm mag}\tilde\Sigma}\psi\right)
\end{equation}
where, now $\tilde\Sigma={\rm diag}\left( 0,0,1,-1\right)$. This
contains the previous identification of the angle $\chi$ as well as a
simultaneous rotation of the transverse $Y$-coordinates. Half of the
fermions are un-twisted and this identification preserves half of the
supersymmetries.  The giant magnon can still be considered a wrapped
state of this orbifold where the identified $Y$-coordinates are not excited.

The gauge theory duals of both of these models are well-known orbifold
 projections of ${\cal N}=4$ theory~\cite{Douglas:1996sw}.  They are obtained by beginning
 with the parent theory, ${\cal N}=4$ super Yang-Mills with gauge
 group $SU(MN)$ and coupling constant $g_{YM}$.  Then, we consider a
 simultaneous $R$-symmetry transformation by a generator of the $Z_M$
 orbifold group and a gauge transform by a constant $SU(MN)$ matrix
 $\gamma={\rm diag}(1,\omega,\omega^2,...,\omega^{M-1})$ where $\omega$
 is the $M$-th root of unity. Each diagonal element of the $MN\times
 MN$-matrix $\gamma$ is multiplied by the $N\times N$ unit matrix. The
 projection throws away all fields which are not invariant under the
 simultaneous transformation. This reduces a typical field which was an $MN\times MN$ matrix
 in the parent theory to $M~$ $N\times N$ blocks embedded in that matrix in the
 orbifold theory.

 For example, consider a field $Z$ of the parent theory which
 is charged under the orbifold group and transforms as $Z\to\omega Z$.
 The orbifold projection reduces it to
 a matrix which obeys   \begin{equation}\label{zorbifold} Z\gamma = \omega \gamma Z\end{equation}
 By similar reasoning, a field $\Phi$ which was neutral in the parent theory commutes with $\gamma$
 once the orbifold projection is imposed,
 \begin{equation}\label{phiorbifold}\Phi\gamma = \gamma\Phi\end{equation}

 Given any single-trace operator of the parent ${\cal N}=4$ theory,
 for example, a single magnon state such as ${\rm Tr} Z^J\Phi$, there
 are a family of $M$ states of the orbifold theory ${\rm
 Tr}\gamma^{m}Z^J\Phi$ with $m=0,1,...,M-1$.  The operator must be
 neutral under the orbifold group transformation in the parent theory.
 To see this: we could insert $1=\gamma^{M-1}\gamma$ into the trace
 and use the commutators such as (\ref{zorbifold}) and
 (\ref{phiorbifold}) and cyclicity of the trace to show that the trace
 of any operator which is not a singlet under the orbifold group must
 vanish.  In our example, if $\Phi$ is neutral, this requires
 quantization of $J$ in units of $M$, $J=kM$, in the state ${\rm
 Tr}\gamma^{m}Z^J\Phi$.  This is the gauge dual of the quantization of
 the momentum $J$ in units of $M\cdot$integers, rather than integers
 after the orbifold projection is imposed in the sigma model,
 discussed above after Eq.~(\ref{J}). In addition, the single-trace
 operator of the parent theory descends to a family of $M$ operators
 which are distinguished an additional quantum number, $m$.  It is
 easy to see that moving the position where $\Phi$ was inserted into
 ${\rm Tr}\gamma^{m}Z^J\Phi$ changes the operator by an overall factor
 of $\omega^m$.  This implies that this trace is already an eigenstate
 of magnon momentum, $p_{\rm mag}=2\pi\frac{m}{M}$.  The integer $m$
 is the gauge theory dual of the wrapping number of the string state
 on the orbifold cycle.

There is a theorem to the effect that, in the planar limit of the
orbifold gauge theory, un-twisted operators (with $m=0$ in the above
examples) have the same correlation functions with each other as
those in the planar parent ${\cal N}=4$ gauge theory -- with the
only difference being a re-scaling of the coupling constant by the
order of the orbifold group \cite{Bershadsky:1998cb}. For this
reason, in the planar limit, the gauge theory resulting from either
of the orbifold projections (\ref{orbifold1}) or (\ref{orbifold2})
is a conformal field theory. In the non-supersymmetric case
(\ref{orbifold1}) non-planar corrections would give a beta-function,
whereas in the ${\cal N}=2$ supersymmetric case (\ref{orbifold2})
the beta function would vanish in the full theory.

On the orbifold, the spectrum of states in the ${\cal N}=4$ magnon
super-multiplet are expected to be split according to the residual
symmetries. In the two cases we considered, the first
(\ref{orbifold1}) has no supersymmetry but has $SU(2)^4\times R^2$
bosonic symmetry.  We would expect that the fermionic states gain
different energies than the bosonic states and that the $SU(2)$
multiplets within the bosonic states also split.  In the other case
(\ref{orbifold2}), there remains ${\cal N}=2$ supersymmetry and the
spectrum should represent the super-algebra $SU(2|1)^2\times R^2$.
The ${\cal N}=4$ magnon supermultiplet becomes
\begin{eqnarray}
{\rm Tr}\gamma^m D_{\mu}ZZ^{kM-1}  \label{ads}\\
\label{s5} {\rm Tr}\gamma^m \Phi Z^{kM} ~~,~~ {\rm Tr}\gamma^m
\bar\Phi Z^{kM}
~~,~~ {\rm Tr}\gamma^m\bar \Psi Z^{kM+1} ~~,~~ {\rm Tr}\gamma^m
\Psi Z^{kM-1}\\
{\rm Tr}\gamma^m  \chi_{1\alpha}Z^{kM}~,~{\rm Tr}\gamma^m
\chi_{3\alpha}Z^{kM-1} ~,~{\rm Tr}\gamma^m \bar\chi^2_{\dot\alpha}
Z^{kM}~,~{\rm Tr}\gamma^m \bar\chi^4_{\dot\alpha} Z^{kM+1}
\label{ferms}\end{eqnarray} 
Here $m$ gives the number of units of magnon momentum $p_{\rm
mag}=\frac{2\pi}{M}m$ and $k$ is the number of units of space-time
momentum $J=kM$.  There are two limits where the operators in the set
(\ref{ads})-(\ref{ferms}) are degenerate and have energies
$\Delta-J=1$: One is when we turn off the 'tHooft coupling
$\lambda=g_{YM}^2MN\to 0$ so that the operators have their classical
conformal dimension. The other is when magnon momentum vanishes,
$m=0$.  In the latter, the ``untwisted operator'' with $m=0$ is known
to have identical correlation functions with the operators in the
parent ${\cal N}=4$ theory and therefore have exact conformal
dimension $\Delta=J+1$. The spectrum away from these limits will
depend on both $\lambda$ and $m$. It would be interesting to check the
splitting of the supermultiplet in perturbative gauge theory, a task
which we reserve for a later publication. In particular, it would be
interesting to study the orbifold Bethe
Ans\"atz~\cite{Beisert:2005he}-\cite{Solovyov:2007pw}.

To conclude, we examine the plane-wave limit of $AdS_5\times S^5$
where the string theory sigma model  is exactly solvable
\cite{Metsaev:2001bj}. We re-define the string coordinates as:
$T=X^+$, $\chi=\frac{1}{\sqrt{\lambda}}X^--X^+$.  This has been
chosen so that $\Delta-J = \frac{1}{i}\left(\frac{\partial}{\partial
T} - \frac{\partial}{\partial\chi}\right)=
\frac{1}{i}\frac{\partial}{\partial X^+}$.  In addition we re-scale
the transverse coordinates $\vec Y\to \vec Y/\lambda^{\frac{1}{4}}$,
$\vec Z\to\vec X/\lambda^{\frac{1}{4}}$. The appropriate plane-wave
limit \cite{Berenstein:2002jq} then takes $\lambda\to \infty$
simultaneously with $\Delta\to\infty$ and $J\to\infty$ with
$\Delta-J$ and $\frac{J}{\sqrt{\lambda}}$ finite. From
(\ref{levelmatching}) we see that the limit should be taken so that
$p_{\rm mag}J$ is finite. This implies that
\begin{equation}\label{scalingofpmag} p_{\rm mag}\sim\frac{1}{\sqrt{\lambda}}\end{equation}
The magnon boundary condition (\ref{chiboundarydoncition})  implies
\begin{equation}\label{identificationofxminus}
X^-(\sigma=\pi)=X^-(\sigma=0) +  p_{\rm mag} \sqrt{\lambda}
\end{equation}
The scaling (\ref{scalingofpmag}) then gives a finite
radius for $X^-$.

 We have already argued that
$J=\frac{1}{i}\frac{\partial}{\partial\chi}=\sqrt{\lambda}\frac{1}{i}\frac{\partial}{\partial
X^-}$ should be quantized in integral units.  In fact, in the magnon
sector, we have argued that the level-matching condition
(\ref{levelmatching}) has a solution only when $p_{\rm
mag}=2\pi\frac{m}{M}$ where $m$ and $M$ are integers and $J$ is
quantized in units of $M$, $J=kM$ with $k$ an integer. To get the
correct scaling of $p_{\rm mag}$ we must therefore take the plane
wave limit by taking $M$ to be large so that
$\frac{M}{\sqrt{\lambda}}$ is held finite.

What is effectively the same limit was discussed in
Ref.~\cite{Mukhi:2002ck} where it was shown to result in a
plane-wave background with a periodically identified null direction,
$X^-\sim X^-+2\pi R^-$ where $R^- = \frac{\sqrt{\lambda}}{M}$. (To
be consistent with (\ref{identificationofxminus})), the integer $m$
which appears in $p_{\rm mag}$ is interpreted is a wrapping number.)
The resulting discrete light-cone quantization of the string on the
plane wave background is a simple generalization of Metsaev's
original solution~\cite{Metsaev:2001bj}. Here, we are interested in
a wrapped sector where $X^-(\sigma=2\pi)=X^-(\sigma=0)+2\pi R^-m$.
In Ref.~\cite{Mukhi:2002ck} the spectrum of the IIB string theory in
this plane wave limit was matched with the appropriate
generalization of the BMN limit of the ${\cal N}=2$ Yang-Mills
theory which is obtained from ${\cal N}=4$ by the orbifold
projection corresponding to (\ref{orbifold2}).  It was also used to
study non-planar corrections~\cite{De Risi:2004bc} and finite-size
corrections at weak coupling~\cite{Astolfi:2006is}.

Together with the limit, we take the light-cone gauge,
$X^+=p^+\tau$.  Periodicity of $X^-$ quantizes $p^+= k/R^-$. We
obtain the sigma model as a free massive worldsheet field theory
\begin{eqnarray}
{\cal L}=-\frac{1}{4\pi} \left\{ \partial_a \vec
Y\cdot\partial^a\vec Y + \partial_a\vec Z\cdot\partial^a\vec Z +
(p^+)^2(Y^2+Z^2)\right\}
\nonumber \\
-\frac{ip^+}{2\pi}
\left(\bar\psi\partial_-\bar\psi+\psi\partial_-\psi +
2ip^+\bar\psi\Pi\psi\right)
\end{eqnarray}
with $\Pi={\rm diag}(1,1,1,1,-1,-1,-1,-1)$.   In this limit, the
magnon parameter $p_{\rm mag}$ does not appear in the Lagrangian or
the mass-shell condition which determines the light-cone
Hamiltonian:
\begin{eqnarray}\label{ppham}
p^-=\frac{1}{p^+}  \sum_{n=-\infty}^\infty \sqrt{n^2+(p^+)^2} \left(
\alpha_{n}^{\alpha_1\dot\alpha_1\dagger}\alpha_{n\alpha_1\dot\alpha_1}+
\alpha_{n}^{\alpha_2\dot\alpha_2\dagger}\alpha_{n\alpha_2\dot\alpha_2}
\right. \nonumber \\ \left. +
\beta_{n}^{\alpha_1\dot\alpha_2\dagger}\beta_{n\alpha_1\dot\alpha_2}+
\beta_{n}^{\alpha_2\dot\alpha_1\dagger}\beta_{n\alpha_2\dot\alpha_1}\right)
\end{eqnarray}
Its only vestige is in the level-matching condition.
\begin{equation}\label{pplevel}
km= \sum_{n=-\infty}^\infty n \left(
\alpha_{n}^{\alpha_1\dot\alpha_1\dagger}\alpha_{n\alpha_1\dot\alpha_1}+
\alpha_{n}^{\alpha_2\dot\alpha_2\dagger}\alpha_{n\alpha_2\dot\alpha_2}+
\beta_{n}^{\alpha_1\dot\alpha_2\dagger}\beta_{n\alpha_1\dot\alpha_2}+
\beta_{n}^{\alpha_2\dot\alpha_1\dagger}\beta_{n\alpha_2\dot\alpha_1}\right)
\end{equation}
where $k$ are the number of units of $J=kM$ and $m$ is the wrapping
number.  The bosonic $\alpha_{n..}$ and fermionic $\beta_{n..}$
oscillators have the non-vanishing brackets
\begin{eqnarray}
[\alpha_{m\alpha_1\dot\alpha_1},\alpha^{\beta_1\dot\beta_1\dagger}_{n}
]= \delta_{mn}
\delta^{\alpha_1}_{\beta_1}\delta^{\dot\alpha_1}_{\dot\beta_1} &,&
\{\beta_{m\alpha_1\dot\alpha_2},\beta^{\beta_1\dot\beta_2\dagger}_{n}
\}= \delta_{mn}
\delta^{\alpha_1}_{\beta_1}\delta^{\dot\alpha_2}_{\dot\beta_2} \\
\left[\alpha_{m\alpha_2\dot\alpha_2},\alpha^{\beta_2
\dot\beta_2\dagger}_{n}\right] =\delta_{mn}
\delta^{\alpha_2}_{\beta_2}\delta^{\dot\alpha_2}_{\dot\beta_2}&,&
\{\beta_{m\alpha_2\dot\alpha_1},\beta^{\beta_2\dot\beta_1\dagger}_{n}
\}= \delta_{mn}
\delta^{\alpha_2}_{\beta_2}\delta^{\dot\alpha_1}_{\dot\beta_1}
\end{eqnarray}
and  bi-spinors of $SO(4)\times SO(4)\sim SU(2)^4$.\footnote{Indices
are raised and lowered with $\epsilon^{\alpha_i\beta_i}$ and
$-\epsilon_{\alpha_i\beta_i}$, respectively, always operating from
the left.}  We confirm in (\ref{pplevel}), which is the plane wave
limit of (\ref{levelmatching}), there is  solution of the level
matching constraint unless $\frac{p_{\rm mag}}{2\pi}J=$integer.
Here, we can think of the null identification as the vestige of the
orbifold identification.

The level-matching condition (\ref{levelmatching}) allows
1-oscillator states and the magnon supermultiplet is the sixteen
states
\begin{equation} \label{planewavesupermultiplet}
\alpha^{\dagger}_{km\alpha_1\dot\alpha_1}|p^+> ~,~
\alpha^{\dagger}_{km\alpha_2\dot\alpha_2}|p^+>~,~
\beta^{\dagger}_{km\alpha_1\dot\alpha_2}|p^+> ~,~
\beta^{\dagger}_{km\alpha_2\dot\alpha_1}|p^+>
\end{equation}
These states are degenerate with spectrum given by \begin{equation}
p^-=\frac{1}{p^+} \sqrt{ (km)^2+(p^+)^2} = \sqrt{ 1+(R^-)^2
m^2}=\sqrt{ 1+\frac{\lambda'}{M^2}  m^2}
\end{equation}
has the form expected from the plane-wave limit of (\ref{spectrum})
when $p_{\rm mag}=2\pi\frac{m}{M}$. Note that in this plane-wave
limit the finite size corrections that occur in (\ref{spectrum2})
vanish due to the limit of small $p_{\rm mag}$.

The degeneracy of the states in (\ref{planewavesupermultiplet}) can
be attributed to an enhancement of the supersymmetry which is well
known to occur in the Penrose limit. One would expect, and we shall
confirm, that the supersymmetry is broken when corrections to the
Penrose limit are taken into account. Before that, we recall that in
Refs.~\cite{Beisert:2005tm},\cite{Beisert:2006qh} Beisert argued
magnon states form a sixteen dimensional short multiplet of an
extended super-algebra $SU(2|2)\times SU(2|2)\times (R^1)^3$ where
the spectrum (\ref{spectrum}) is the shortening condition. The
superalgebra $SU(2|2)$ has generators ${\cal
R}^{\alpha_1}_{~\beta_1}$ and ${\cal
L}^{\dot\alpha_2}_{~\dot\beta_2}$ of $SU(2)\times SU(2)$,
supercharges ${\cal Q}^{\dot\alpha_2}_{~\alpha_1}$ and ${\cal
S}_{~\dot\alpha_2}^{\alpha_1}$ and the algebra

\begin{eqnarray}
\left[ {\cal R}^{\alpha_1}_{~\beta_1},{\cal
J}^{\gamma_1}\right]=\delta^{\gamma_1}_{\beta_1} {\cal
J}^{\alpha_1}-\frac{1}{2}\delta^{\alpha_1}_{\beta_1} {\cal
J}^{\gamma_1} &,& \left[ {\cal L}^{\dot\alpha_2}_{~\dot\beta_2},{\cal
J}^{\dot\gamma_2}\right]=\delta^{\dot\gamma_2}_{\dot\beta_2} {\cal
J}^{\dot\alpha_2}-\frac{1}{2}\delta^{\dot\alpha_2}_{\dot\beta_2} {\cal
J}^{\dot\gamma_2}
\nonumber \\
\left\{{\cal Q}^{\dot\alpha_2}_{~\alpha_1},{\cal
S}^{\beta_1}_{~\dot\beta_2}\right\}&=&\delta^{\beta_1}_{\alpha_1}{\cal
L}^{\dot\alpha_2}_{~\dot\beta_2}+\delta^{\dot\alpha_2}_{\dot\beta_2} {\cal
R}^{\beta_1}_{~\alpha_1}+\delta^{\beta_1}_{\alpha_1}\delta^{\dot\alpha_2}_{\dot\beta_2}
{\cal C}
\nonumber \\
\left\{ {\cal Q}^{\dot\alpha_2}_{~\alpha_1},{\cal
Q}^{\dot\beta_2}_{~\beta_1}\right\}=\epsilon^{\dot\alpha_2\dot\beta_2}
\epsilon_{\alpha_1\beta_1}{\cal
P} ~~&,&~~ \left\{ {\cal S}_{~\dot\alpha_2}^{\alpha_1},{\cal
S}_{~\dot\beta_2}^{\beta_1}\right\}=\epsilon_{\dot\alpha_2\dot\beta_2}
\epsilon^{\alpha_1\beta_1}{\cal K}
\nonumber \end{eqnarray}
${\cal J}^{...}$ represents any generator with the appropriate index, ${\cal
K}$, ${\cal P}$ and ${\cal C}$ are central charges. In our
application, ${\cal C}=\Delta-J = p^-$ and
\begin{eqnarray}
{\cal R}^{\alpha_1}_{~\beta_1}
&=&\sum_n\left\{\alpha^{\dagger\alpha_1\dot\gamma}_n\alpha_{n\beta_1\dot\gamma_1}
+\beta_n^{\dagger\alpha_1\gamma_2}\beta_{\beta_1\gamma_2}\right\} -
\frac{1}{2} \delta^{\alpha_1}_{~\beta_1}\sum_n
\left\{\alpha^{\dagger\gamma_1\dot\gamma_1}_n\alpha_{n\gamma_1\dot\gamma_1}
+\beta_n^{\dagger\gamma_1\gamma_2}\beta_{\gamma_1\gamma_2}\right\}
\nonumber \\
{\cal L}^{\dot\alpha_2}_{~\dot\beta_2}&=&
\sum_n\left\{\alpha^{\dagger\gamma_2\dot\alpha_2}_n\alpha_{n\gamma_2\dot\beta_2}
+\beta_n^{\dagger\dot\alpha_2\dot\gamma_1}\beta_{\dot\gamma_1\dot\beta_2}\right\}
- \frac{1}{2} \delta^{\dot\alpha_2}_{~\dot\beta_2}\sum_n
\left\{\alpha^{\dagger\gamma_2\dot\gamma_2}_n\alpha_{n\gamma_2\dot\gamma_2}
+\beta_n^{\dagger\dot\gamma_1\dot\gamma_2}\beta_{\dot\gamma_1\dot\gamma_2}\right\}
\nonumber \\
{\cal Q}_{~\alpha_1}^{\dot\beta_2}&=&\frac{\bar\eta}{\sqrt{8
p^+}}\sum_n\left\{-e(n)\sqrt{\omega_n+p^+}
\alpha^{\dagger}_{n\alpha_1\dot\gamma_1}\beta_{n}^{\dot\gamma_1\dot\beta_2}
+i\sqrt{\omega_n-p^+}\alpha_{n\alpha_1\dot\gamma_1}\beta_{n}^{\dagger\dot\gamma_1\dot\beta_2}
 \right. \nonumber \\  &~&\left. ~~~~~~~~~~~~-i\sqrt{\omega_n-p^+}\beta^{\dagger}_{n\alpha_1\gamma_2}\alpha_{n}^{\gamma_2\dot\beta_2}
+ e(n)\sqrt{\omega_n+p^+}
\beta_{n\alpha_1\gamma_2}\alpha^{\dagger\gamma_2\dot\beta_2}_n
 \right\}\nonumber\\
{\cal S}^{\alpha_1}_{~\dot\beta_2}&=&\frac{\bar\eta}{\sqrt{8 p^+}}
\sum_n\left\{  \sqrt{\omega_n-p^+}
\alpha^{\dagger\alpha_1\dot\gamma_1}_n\beta_{n\dot\gamma_1\dot\beta_2}
-i e(n)
\sqrt{\omega_n+p^+}\alpha^{\alpha_1\dot\gamma_1}_n\beta^{\dagger}_{n\dot\gamma_1\dot\beta_2}
+ \right. \nonumber \\  &~&\left. ~~~~~~~~~~~~ +i e(n)
\sqrt{\omega_n+p^+}\beta^{\dagger\alpha_1\gamma_2}_{n}\alpha_{n\gamma_2\dot\beta_2}
-
\sqrt{\omega_n-p^+}\beta^{\alpha_1\gamma_2}_{n}\alpha^{\dagger}_{n\gamma_2\dot\beta_2}
 \right\}\nonumber \\
\end{eqnarray}
where  $\omega_n=\sqrt{(p^+)^2+n^2}$ and $e(n)=\frac{n}{|n|}$.  We
have used Metsaev's \cite{Metsaev:2001bj} conventions for the
supercharges (those called $Q^-$ and $\bar Q^-$) and notation for
oscillators as summarized, for example, in Ref.~\cite{Young:2007pk}.
Computing their algebra, we find that the plane wave background
supercharges indeed satisfy Beisert's extended superalgebra with the
central extensions set to the plane-wave limits of those found by
Beisert~\cite{Beisert:2005tm} \begin{equation}{\cal
P}=-i\frac{\sqrt{\lambda}p_{\rm mag}}{4\pi}\leftarrow
\frac{\sqrt{\lambda}}{4\pi}\left(e^{-ip_{\rm mag}}-1\right) ~,~
{\cal K}=i\frac{\sqrt{\lambda}p_{\rm mag}}{4\pi}\leftarrow
\frac{\sqrt{\lambda}}{4\pi}\left(e^{ip_{\rm mag}}-1\right)
\label{extension}\end{equation} The existence of the central
extension follows directly from the fact that the unextended algebra
closes up to the level matching condition and the level-matching
condition (\ref{levelmatching}) contains the term with $km=
\frac{1}{2\pi}2\pi\frac{m}{M}\cdot kM=\frac{1}{2\pi}p_{\rm mag}J$.

A derivation of Beisert's superalgebra in the context of the
$AdS_3\times S^5$ sigma model was first given in
Ref.~\cite{Arutyunov:2006ak} and developed in
Ref.~\cite{Arutyunov:2006yd}.  They worked with the un-orbifolded
theory by ``relaxing'' the level-matching condition.  Then, there is a
central charge in the superalgebra which depends on the level
miss-match. The idea is that, once the resulting algebraic structure
is used to study magnon and multi-magnon states, the level-matching
condition should be re-imposed so as to get a physical state of the
string theory. They work in the ``magnon limit'', where $J\to\infty$,
but magnon momentum is not necessarily small (in our case it relaxes
the plane-wave limit by taking $M$ not necessarily large). They obtain
the full central extension, rather than the form linearized in $p_{\rm
mag}$ that we have found in (\ref{extension}).  In their work, they
use a generalized light-cone gauge $x^+=\tau=(1-a)T+a\chi$,
$x_-=\chi-T$ with $a$ a parameter.  They also use the identification,
$x_-(\tau,\sigma=2\pi)-x_-(\tau,\sigma=0) =p_{\rm ws}$ with $p_{\rm
ws}$ an eigenvalue of the level operator and $x_+=\tau$ trivially
periodic in $\sigma$. For the variables in (\ref{sigmamodelaction}),
this amounts to using the boundary condition
$\chi(\tau,\sigma=2\pi)-\chi(\tau,\sigma=0)= -(1-a)p_{\rm ws}$ and
$T(\tau,\sigma=2\pi)-T(\tau,0)=ap_{\rm ws}$ which is different from
the one which we use when $a\neq0$ (they primarily use
$a=\frac{1}{2}$) - where $T(\tau,\sigma=2\pi)=T(\tau,\sigma=0)$ and
$\chi(\tau,\sigma=2\pi)-\chi(\tau,\sigma=0)=p_{\rm mag}$.  This makes
no difference at infinite $J$ where the effect of $a$ is diluted by
scaling.  However, it matters at finite size. In fact, the same gauge
fixing was used in Ref.~\cite{Arutyunov:2006gs} and the $a$-dependence
of the one-magnon spectrum found there (away from the infinite $J$
limit) can be attributed to this $a$-dependence of boundary
conditions, rather than the gauge variance which is claimed there.

To see how the spectrum will be split in the near plane-wave limit, we
must include corrections to the Lagrangian and the Virasoro
constraints that are of order $\frac{1}{\sqrt{\lambda}}$.  A
systematic scheme for including these corrections in the usual $p_{\rm
mag}=0$ sector are outlined in the series of
papers~\cite{Callan:2004ev}-\cite{Callan:2003xr} and nicely summarized
in Ref.~\cite{McLoughlin:2005dh}. There they find that the corrections
terms to the Hamiltonian add normal ordered terms which are quartic in
oscillators.  They also adjust the gauge by adjusting the worldsheet
metric in such a way that the level-matching condition remains
unmodified.  We have shown, and will present elsewhere, that the
modification of at procedure in the magnon sector are minimal.  The
corrections to the free field theory light-cone Hamiltonian are of two
types, quartic normal ordered pieces from near-plane-wave limit
corrections to the sigma model identical in form to those found in
Refs.~\cite{Callan:2004ev}-\cite{McLoughlin:2005dh} and terms such as
the last one in Eq.~(\ref{shiftedlagrangian}) which arise from the
orbifolding.

To leading order in perturbation theory, the normal ordered quartic
interaction Hamiltonian cannot shift the spectrum of 1-oscillator
states.  Furthermore, none of the extra terms displayed in
Eq.~(\ref{shiftedlagrangian}) contribute in the leading order in
$1/\sqrt{\lambda}$.   However, recall that, to preserve some
supersymmetry, the orbifold identification (\ref{orbifold2}) that we
have been discussing also acts on the transverse direction and this
action must also be taken into account.  This generates simple
correction terms in the Hamiltonian to order
$\frac{1}{\sqrt{\lambda}}$.   The relevant part of the interaction
Hamiltonian is
\begin{equation}
H_{\rm int}= i\frac{p_{\rm mag}}{2\pi}\frac{1}{2\pi}\int_0^2\pi
d\sigma\left( Y_{1_1\dot 2_1}Y_{2_1\dot 1_1}' +ip^+\left(
\psi\tilde\Sigma\psi+\bar\psi\tilde\Sigma\bar\psi\right)\right)
\end{equation}


With this orbifold identification exactly half of the supersymmetries
are preserved in the near plane-wave limit. Specifically, out of the
16 supersymmetries $ {\cal Q}_{\a_1}^{\dot \a_2}, {\cal S}_{\dot
\a_2}^{\a_1}$ only ${\cal S}^{\a_1}_{\dot 1_2} / {\cal Q}_{\a_1}^{\dot 1_2}$ and ${\cal S}^{2_2}_{\dot \a_1} / {\cal Q}_{2_2}^{\dot \a_1} $ survive. This leads to a splitting of the energies of the
single impurity states.

The original multiplet had 16 states (8 bosons - $\a^\dagger_{\a_1
\dot \a_1} |0>, \a^\dagger_{\a_2 \dot \a_2} |0>$ and 8 fermions -
$\b^\dagger_{\a_1 \a_2} |0>, \b^\dagger_{\dot \a_1 \dot \a_2}
|0>$). In the near plane-wave, it breaks up into 4 super-multiplets of
the residual superalgebra: one with 9 elements (5 bosons and 4
fermions) and two with 3 elements (2 fermions and a boson in each) and
one boson singlet.

The following table illustrates the breaking of the original
super-multiplet: 

\be
\scriptsize
\label{multiplets}
\matrix{&& {\cal S}^{2_1}_{\dot 1_2} / {\cal Q}_{2_1}^{\dot 1_2} &&  {\cal Q}_{1_1}^{\dot 1_2}/{\cal S}^{1_1}_{\dot 1_2}&&
{\cal S}^{1_1}_{\dot 2_2} / {\cal Q}_{1_1}^{\dot 2_2}&&{\cal Q}_{2_1}^{\dot 2_2}/{\cal S}^{2_1}_{\dot 2_2}\cr
&\a_{2_1 \dot 2_1}^\dagger|0>&-&\b_{\dot 2_1 \dot 1_2}^\dagger|0>&-&\a_{1_1 \dot 2_1}^\dagger|0>&&\b_{\dot 2_1 \dot 2_2}^\dagger|0>&\cr
{\cal Q}_{2_2}^{\dot 2_1}/{\cal S}^{2_2}_{\dot 2_1}&|&&|&&|&&|&\cr
&\b_{2_1 2_2}^\dagger|0>&-&\a_{2_2 \dot 1_2}^\dagger|0>&-&\b_{1_1 2_2}^\dagger|0>&&\a_{2_2 \dot 2_2}^\dagger|0>&\cr
{\cal S}^{2_2}_{\dot 1_1} / {\cal Q}_{2_2}^{\dot 1_1}&|&&|&&|&&|&\cr
&\a_{2_1 \dot 1_1}^\dagger|0>&-&\b_{\dot 1_1 \dot 1_2}^\dagger|0>&-&\a_{1_1 \dot 1_1}^\dagger|0>&&\b_{\dot 1_1 \dot 2_2}^\dagger|0>&\cr
{\cal Q}_{1_2}^{\dot 1_1}/{\cal S}^{1_2}_{\dot 1_1}&&&&&&&&\cr
&\b_{2_1 1_2}^\dagger|0>&-&\a_{1_2 \dot 1_2}^\dagger|0>&-&\b_{1_1 1_2}^\dagger|0>&&\a_{1_2 \dot 2_2}^\dagger|0>&\cr
{\cal S}^{1_2}_{\dot 2_1} / {\cal Q}_{2_1}^{\dot 1_2}&&&&&&&&\cr}
\ee

Here, columns and rows with dashes represent the surviving supersymmetry transformations:
 ${\cal S}^{\a_1}_{\dot 1_2} / {\cal Q}_{\a_1}^{\dot 1_2}$ and ${\cal S}^{2_2}_{\dot \a_1} / {\cal Q}_{2_2}^{\dot \a_1} $. Columns and rows without dashes represent the broken supersymmetries:  ${\cal S}^{\a_1}_{\dot 2_2} / {\cal Q}_{\a_1}^{\dot 2_2}$ and ${\cal S}^{1_2}_{\dot \a_1} / {\cal Q}_{1_2}^{\dot \a_1} $.

The energy degeneracy of the original multiplet is likewise broken by
the interaction Hamiltonian in the near plane-wave limit.  One of the
triplets gets positive energy shift, its energy becoming:
$$\sqrt{ 1+\lambda\frac{m^2}{M^2} }+\frac{1}{2\sqrt{\lambda}}~
\frac{\lambda\frac{m^2}{M^2} }{ \sqrt{ 1+\lambda\frac{m^2}{M^2} }}$$
Other triplet gets equal but negative energy shift:
 $$\sqrt{ 1+\lambda\frac{m^2}{M^2} }-\frac{1}{2\sqrt{\lambda}}~ \frac{
\lambda\frac{m^2}{M^2} }{ \sqrt{ 1+\lambda\frac{m^2}{M^2} }}$$
Singlet and a 9-multiplet are annihilated by the interaction
Hamiltonian and thus retain the energy of the original multiplet:
 $$\sqrt{1+\lambda\frac{m^2}{M^2} }$$

In conclusion, we have made an number of observations about the
giant magnon solution of string theory.  We observed that the
previously noted resemblance of the magnon to a wrapped string on a
$Z_M$  orbifold of $AdS_5\times S^5$ seems to be the only solution
of the Virasoro constraints in the string sigma-model. We argued
that this point of view is consistent with AdS/CFT duality as single
magnons are physical states of the orbifold projections of ${\cal
N}=4$ supersymmetric Yang-Mills theory.  We also argued that this
point of view is consistent with the plane wave limit, where the
sigma model is solvable.  In that limit, the orbifold identification
appears as a periodic identification of the null coordinate and the
magnon is a wrapped string.  There, we can see explicitly how the
wrapping modifies the supersymmetry algebra and is consistent with
the magnon spectrum. The ${\cal N}=2$ supersymmetry of the orbifold
is enhanced to ${\cal N}=4$ supersymmetry in the plane wave limit,
so that the full sixteen dimensional magnon supermultiplet appears
there.   We end with a question.  We have shown that the
supersymmetry is broken again by near-plane wave limit corrections
to the sigma model by showing that the energies of the magnon
multiplet are split. However, there is another limit, the ``magnon
limit'' which is similar to the plane wave in that $\lambda$ and $J$
are taken to infinity but it differs in that $p_{\rm mag}$ remains
of order one, rather than scaling to zero. It would be interesting
to understand whether the supersymmetry is also enhanced in this
limit so that the orbifold quantization of the infinite volume limit
has more supersymmetry than the orbifold itself.


\begin{thebibliography}{0}

\bibitem{Maldacena:1997re}
  J.~M.~Maldacena,
  ``The large N limit of superconformal field theories and supergravity,''
  Adv.\ Theor.\ Math.\ Phys.\  {\bf 2}, 231 (1998)
  [Int.\ J.\ Theor.\ Phys.\  {\bf 38}, 1113 (1999)]
  [arXiv:hep-th/9711200].

\bibitem{Gubser:1998bc}
  S.~S.~Gubser, I.~R.~Klebanov and A.~M.~Polyakov,
  ``Gauge theory correlators from non-critical string theory,''
  Phys.\ Lett.\  B {\bf 428}, 105 (1998)
  [arXiv:hep-th/9802109].

\bibitem{Witten:1998qj}
  E.~Witten,
  ``Anti-de Sitter space and holography,''
  Adv.\ Theor.\ Math.\ Phys.\  {\bf 2}, 253 (1998)
  [arXiv:hep-th/9802150].

\bibitem{Minahan:2002ve}
  J.~A.~Minahan and K.~Zarembo,
  ``The Bethe-ansatz for N = 4 super Yang-Mills,''
  JHEP {\bf 0303}, 013 (2003)
  [arXiv:hep-th/0212208].

\bibitem{Beisert:2003tq}
  N.~Beisert, C.~Kristjansen and M.~Staudacher,
  ``The dilatation operator of N = 4 super Yang-Mills theory,''
  Nucl.\ Phys.\  B {\bf 664}, 131 (2003)
  [arXiv:hep-th/0303060].

\bibitem{Beisert:2003yb}
  N.~Beisert and M.~Staudacher,
  ``The N = 4 SYM integrable super spin chain,''
  Nucl.\ Phys.\  B {\bf 670}, 439 (2003)
  [arXiv:hep-th/0307042].

\bibitem{Arutyunov:2004vx}
  G.~Arutyunov, S.~Frolov and M.~Staudacher,
  ``Bethe ansatz for quantum strings,''
  JHEP {\bf 0410}, 016 (2004)
  [arXiv:hep-th/0406256].


\bibitem{Staudacher:2004tk}
  M.~Staudacher,
  ``The factorized S-matrix of CFT/AdS,''
  JHEP {\bf 0505} (2005) 054
  [arXiv:hep-th/0412188].



\bibitem{Beisert:2005tm}
  N.~Beisert,
  ``The su(2|2) dynamic S-matrix,''
  arXiv:hep-th/0511082.

\bibitem{Beisert:2006qh}
  N.~Beisert,
  ``The Analytic Bethe Ansatz for a Chain with Centrally Extended su(2|2)
  Symmetry,''
  J.\ Stat.\ Mech.\  {\bf 0701}, P017 (2007)
  [arXiv:nlin/0610017].

\bibitem{Janik:2006dc}
  R.~A.~Janik,
  ``The AdS(5) x S**5 superstring worldsheet S-matrix and crossing  symmetry,''
  Phys.\ Rev.\  D {\bf 73}, 086006 (2006)
  [arXiv:hep-th/0603038].





\bibitem{Hernandez:2006tk}
  R.~Hernandez and E.~Lopez,
  ``Quantum corrections to the string Bethe ansatz,''
  JHEP {\bf 0607}, 004 (2006)
  [arXiv:hep-th/0603204].

\bibitem{Beisert:2006zy}
  N.~Beisert,
  ``On the scattering phase for AdS(5) x S**5 strings,''
  Mod.\ Phys.\ Lett.\  A {\bf 22}, 415 (2007)
  [arXiv:hep-th/0606214].

\bibitem{Beisert:2006ib}
  N.~Beisert, R.~Hernandez and E.~Lopez,
  ``A crossing-symmetric phase for AdS(5) x S**5 strings,''
  JHEP {\bf 0611}, 070 (2006)
  [arXiv:hep-th/0609044].

\bibitem{Beisert:2006ez}
  N.~Beisert, B.~Eden and M.~Staudacher,
  ``Transcendentality and crossing,''
  J.\ Stat.\ Mech.\  {\bf 0701} (2007) P021
  [arXiv:hep-th/0610251].




\bibitem{Freyhult:2006vr}
  L.~Freyhult and C.~Kristjansen,
  ``A universality test of the quantum string Bethe ansatz,''
  Phys.\ Lett.\  B {\bf 638}, 258 (2006)
  [arXiv:hep-th/0604069].


\bibitem{Benna:2006nd}
  M.~K.~Benna, S.~Benvenuti, I.~R.~Klebanov and A.~Scardicchio,
  ``A test of the AdS/CFT correspondence using high-spin operators,''
  Phys.\ Rev.\ Lett.\  {\bf 98}, 131603 (2007)
  [arXiv:hep-th/0611135].




\bibitem{Fiamberti:2007rj}
  F.~Fiamberti, A.~Santambrogio, C.~Sieg and D.~Zanon,
  ``Wrapping at four loops in N=4 SYM,''
  arXiv:0712.3522 [hep-th].

\bibitem{Keeler:2008ce}
  C.~A.~Keeler and N.~Mann,
  ``Wrapping Interactions and the Konishi Operator,''
  arXiv:0801.1661 [hep-th].


\bibitem{Ambjorn:2005wa}
  J.~Ambjorn, R.~A.~Janik and C.~Kristjansen,
  ``Wrapping interactions and a new source of corrections to the spin-chain  /
  string duality,''
  Nucl.\ Phys.\  B {\bf 736}, 288 (2006)
  [arXiv:hep-th/0510171].

\bibitem{Kotikov:2007cy}
  A.~V.~Kotikov, L.~N.~Lipatov, A.~Rej, M.~Staudacher and V.~N.~Velizhanin,
  ``Dressing and Wrapping,''
  J.\ Stat.\ Mech.\  {\bf 0710}, P10003 (2007)
  [arXiv:0704.3586 [hep-th]].

\bibitem{Janik:2007wt}
  R.~A.~Janik and T.~Lukowski,
  ``Wrapping interactions at strong coupling -- the giant magnon,''
  Phys.\ Rev.\  D {\bf 76}, 126008 (2007)
  [arXiv:0708.2208 [hep-th]].




\bibitem{Hofman:2006xt}
  D.~M.~Hofman and J.~M.~Maldacena,
  ``Giant magnons,''
  J.\ Phys.\ A  {\bf 39}, 13095 (2006)
  [arXiv:hep-th/0604135].


\bibitem{Arutyunov:2006gs}
  G.~Arutyunov, S.~Frolov and M.~Zamaklar,
  ``Finite-size effects from giant magnons,''
  Nucl.\ Phys.\  B {\bf 778}, 1 (2007)
  [arXiv:hep-th/0606126].

\bibitem{Astolfi:2007uz}
  D.~Astolfi, V.~Forini, G.~Grignani and G.~W.~Semenoff,
  ``Gauge invariant finite size spectrum of the giant magnon,''
  Phys.\ Lett.\  B {\bf 651}, 329 (2007)
  [arXiv:hep-th/0702043].



\bibitem{Ishizeki:2007we}
  R.~Ishizeki and M.~Kruczenski,
  ``Single spike solutions for strings on S2 and S3,''
  Phys.\ Rev.\  D {\bf 76}, 126006 (2007)
  [arXiv:0705.2429 [hep-th]].


\bibitem{Hatsuda:2008gd}
  Y.~Hatsuda and R.~Suzuki,
  ``Finite-Size Effects for Dyonic Giant Magnons,''
  arXiv:0801.0747 [hep-th].


\bibitem{Minahan:2008re}
  J.~A.~Minahan and O.~Ohlsson Sax,
  ``Finite size effects for giant magnons on physical strings,''
  arXiv:0801.2064 [hep-th].

\bibitem{Hayashi:2007bq}
  H.~Hayashi, K.~Okamura, R.~Suzuki and B.~Vicedo,
  ``Large Winding Sector of AdS/CFT,''
  JHEP {\bf 0711}, 033 (2007)
  [arXiv:0709.4033 [hep-th]].

\bibitem{Gromov:2008ie}
  N.~Gromov, S.~Schafer-Nameki and P.~Vieira,
  ``Quantum Wrapped Giant Magnon,''
  arXiv:0801.3671 [hep-th].

\bibitem{Klose:2008rx}
  T.~Klose and T.~McLoughlin,
  ``Interacting finite-size magnons,''
  arXiv:0803.2324 [hep-th].


\bibitem{Polchinski:1998rq}
  J.~Polchinski,
  ``String theory. Vol. 1: An introduction to the bosonic string,''
{\it  Cambridge, UK: Univ. Pr. (1998) 402 p}.

\bibitem{Douglas:1996sw}
  M.~R.~Douglas and G.~W.~Moore,
  ``D-branes, Quivers, and ALE Instantons,''
  arXiv:hep-th/9603167.


\bibitem{Alday:2005ww}
  L.~F.~Alday, G.~Arutyunov and S.~Frolov,
  ``Green-Schwarz strings in TsT-transformed backgrounds,''
  JHEP {\bf 0606}, 018 (2006)
  [arXiv:hep-th/0512253].

\bibitem{Bershadsky:1998cb}
  M.~Bershadsky and A.~Johansen,
  ``Large N limit of orbifold field theories,''
  Nucl.\ Phys.\  B {\bf 536}, 141 (1998)
  [arXiv:hep-th/9803249].

\bibitem{Beisert:2005he}
  N.~Beisert and R.~Roiban,
  ``The Bethe ansatz for Z(S) orbifolds of N = 4 super Yang-Mills theory,''
  JHEP {\bf 0511}, 037 (2005)
  [arXiv:hep-th/0510209].

\bibitem{Solovyov:2007pw}
  A.~Solovyov,
  ``Bethe Ansatz Equations for General Orbifolds of N=4 SYM,''
  arXiv:0711.1697 [hep-th].

\bibitem{Metsaev:2001bj}
  R.~R.~Metsaev,
  ``Type IIB Green-Schwarz superstring in plane wave Ramond-Ramond
  background,''
  Nucl.\ Phys.\  B {\bf 625}, 70 (2002)
  [arXiv:hep-th/0112044].



\bibitem{Berenstein:2002jq}
  D.~E.~Berenstein, J.~M.~Maldacena and H.~S.~Nastase,
  ``Strings in flat space and pp waves from N = 4 super Yang Mills,''
  JHEP {\bf 0204}, 013 (2002)
  [arXiv:hep-th/0202021].




\bibitem{Mukhi:2002ck}
  S.~Mukhi, M.~Rangamani and E.~P.~Verlinde,
  ``Strings from quivers, membranes from moose,''
  JHEP {\bf 0205}, 023 (2002)
  [arXiv:hep-th/0204147].



\bibitem{De Risi:2004bc}
  G.~De Risi, G.~Grignani, M.~Orselli and G.~W.~Semenoff,
  ``DLCQ string spectrum from N = 2 SYM theory,''
  JHEP {\bf 0411} (2004) 053
  [arXiv:hep-th/0409315].

\bibitem{Astolfi:2006is}
  D.~Astolfi, V.~Forini, G.~Grignani and G.~W.~Semenoff,
  ``Finite size corrections and integrability of N = 2 SYM and DLCQ strings on
  a pp-wave,''
  JHEP {\bf 0609}, 056 (2006)
  [arXiv:hep-th/0606193].

\bibitem{Young:2007pk}
  D.~Young,
  ``The AdS/CFT Correspondence: Classical, Quantum, and Thermodynamical
  Aspects,''
  arXiv:0706.3751 [hep-th].


\bibitem{Arutyunov:2006ak}
  G.~Arutyunov, S.~Frolov, J.~Plefka and M.~Zamaklar,
  ``The off-shell symmetry algebra of the light-cone AdS(5) x S**5
  superstring,''
  J.\ Phys.\ A  {\bf 40}, 3583 (2007)
  [arXiv:hep-th/0609157].

\bibitem{Arutyunov:2006yd}
  G.~Arutyunov, S.~Frolov and M.~Zamaklar,
  ``The Zamolodchikov-Faddeev algebra for AdS(5) x S**5 superstring,''
  JHEP {\bf 0704}, 002 (2007)
  [arXiv:hep-th/0612229].


\bibitem{Callan:2004ev}
  C.~G.~.~Callan, T.~McLoughlin and I.~Swanson,
  ``Higher impurity AdS/CFT correspondence in the near-BMN limit,''
  Nucl.\ Phys.\  B {\bf 700}, 271 (2004)
  [arXiv:hep-th/0405153].

\bibitem{Callan:2004uv}
  C.~G.~.~Callan, T.~McLoughlin and I.~Swanson,
  ``Holography beyond the Penrose limit,''
  Nucl.\ Phys.\  B {\bf 694}, 115 (2004)
  [arXiv:hep-th/0404007].

\bibitem{Callan:2003xr}
  C.~G.~.~Callan, H.~K.~Lee, T.~McLoughlin, J.~H.~Schwarz, I.~Swanson and X.~Wu,
  ``Quantizing string theory in AdS(5) x S**5: Beyond the pp-wave,''
  Nucl.\ Phys.\  B {\bf 673}, 3 (2003)
  [arXiv:hep-th/0307032].

\bibitem{McLoughlin:2005dh}
  T.~McLoughlin,
  ``The near-Penrose limit of AdS/CFT,''
  . UMI-31-97357, 2005. Caltech Ph.D. Thesis



\end{thebibliography}
\end{document}